\documentclass[nonacm, manuscript, screen]{acmart}

\newcommand{\hl}[1]{#1}

\usepackage[normalem]{ulem}
\newcommand\redsout{\bgroup\markoverwith{\textcolor{red}{\rule[0.5ex]{2pt}{0.4pt}}}\ULon}

\AtBeginDocument{%
  \providecommand\BibTeX{{%
    \normalfont B\kern-0.5em{\scshape i\kern-0.25em b}\kern-0.8em\TeX}}}

\setcopyright{acmlicensed}
\copyrightyear{2018}
\acmYear{2018}
\acmDOI{XXXXXXX.XXXXXXX}

\acmConference[Conference acronym 'XX]{Make sure to enter the correct
  conference title from your rights confirmation emai}{June 03--05,
  2018}{Woodstock, NY}
\acmBooktitle{Woodstock '18: ACM Symposium on Neural Gaze Detection,
 June 03--05, 2018, Woodstock, NY} 
\acmISBN{978-1-4503-XXXX-X/18/06}

\begin{document}

\title[Advancing Multi-Entity VR Simulators]{Advancing VR Simulators for Autonomous Vehicle–Pedestrian Interactions: A Focus on Multi-Entity Scenarios}

\author{Tram Thi Minh Tran}
\email{tram.tran@sydney.edu.au}
\orcid{0000-0002-4958-2465}
\affiliation{
  \institution{Design Lab, Sydney School of Architecture, Design and Planning, The University of Sydney}
  \city{Sydney}
  \state{NSW}
  \country{Australia}
}

\author{Callum Parker}
\email{callum.parker@sydney.edu.au}
\orcid{0000-0002-2173-9213}
\affiliation{
  \institution{Design Lab, Sydney School of Architecture, Design and Planning, The University of Sydney}
  \city{Sydney}
  \state{NSW}
  \country{Australia}
}

\renewcommand{\shortauthors}{Tran and Parker}

\begin{abstract}
Recent research has increasingly focused on how autonomous vehicles (AVs) communicate with pedestrians in complex traffic situations involving multiple vehicles and pedestrians. VR is emerging as an effective tool to simulate these multi-entity scenarios, offering a safe and controlled study environment. Despite its growing use, there is a lack of thorough investigation into the effectiveness of these VR simulations, leaving a notable gap in documented insights and lessons. This research undertook a retrospective analysis of two distinct VR-based studies: one focusing on multiple AV scenarios (N=32) and the other on multiple pedestrian scenarios (N=25). Central to our examination are the participants' sense of presence and their crossing behaviour. The findings highlighted key factors that either enhance or diminish the sense of presence in each simulation, providing considerations for future improvements. Furthermore, they underscore the influence of controlled scenarios on crossing behaviour and interactions with AVs, advocating for the exploration of more natural and interactive simulations that better reflect real-world AV and pedestrian dynamics. Through this study, we set a groundwork for advancing VR simulators to study complex interactions between AVs and pedestrians.
\end{abstract}


\maketitle

\section{Introduction}

As the prospect of autonomous vehicles (AVs) integrating into urban traffic becomes more imminent, the importance of understanding their interactions with pedestrians has intensified~\cite{rasouli2019autonomous, moore2019case}. This need has catalysed the growth of pedestrian-in-the-loop virtual reality (VR) simulations, a unique platform to evaluate AV–pedestrian interactions in an immersive, safe, and controlled setting~\cite{tran2021review, nascimento2019role}. These simulations are particularly advantageous for prototyping external Human–Machine Interfaces (eHMIs)~\cite{rouchitsas2019external, dey2020taming}—systems that communicate a vehicle's intentions or operational status to pedestrians. They allow for rapid and cost-effective iterations of eHMI designs, especially those utilising emerging technologies like smart infrastructures~\cite{hollander2022take} and wearable augmented reality~\cite{tran2023simulating, tabone2023augmented}.

In recent years, research on AV external communication with pedestrians has expanded to encompass complex traffic scenarios involving multiple vehicles and pedestrians, referred to as multi-entity scenarios. This development seeks to explore how eHMIs could scale up from individual one-on-one scenarios to broader traffic contexts~\cite{tran2023scoping, tran2021review}. However, \hl{the number of} current VR simulations for these purposes is limited, as highlighted by scoping reviews from \citet{tran2021review} and \citet{colley2020unveiling}. There is also a noticeable lack of retrospective analyses of their effectiveness, resulting in a scarcity of documented insights. Retrospective analysis in this context refers to examining data related to pedestrian perceptions and behaviour in the VR simulations to assess their effectiveness in replicating real-world AV–pedestrian interactions. Such analyses are crucial as they provide practical considerations vital for the ongoing improvement of VR simulation practices within the research community. This gap in knowledge hinders our understanding of how to effectively simulate complex interactions between AVs and pedestrians in multi-entity scenarios.

To address this gap, our research undertook a retrospective analysis of two VR studies focused on multi-vehicle and multi-pedestrian scenarios. Utilising presence questionnaires and semi-structured interviews from these studies, we aimed to explore a central research question: \textbf{How do participants perceive their sense of presence and crossing behaviour in multi-entity VR simulations?} Understanding participants' sense of presence—a concept defined as the feeling of `being there' in a virtual environment~\cite{steuer1995defining, lombard1997heart}—is fundamental to assessing the effectiveness of VR simulations. Insights into this aspect can reveal the degree of immersion and realism experienced by participants, which in turn influences their behaviour in the virtual setting.

The contribution of our paper is twofold: Firstly, we provide insights into the efficacy of VR simulations in representing complex AV–pedestrian interactions, offering a balanced view of the current capabilities and limitations of these methods. Secondly, we present practical considerations and a future roadmap for the design and implementation of more effective VR pedestrian simulators. These considerations are specifically crafted to enhance participants' sense of presence and to foster more natural interactions between AVs and pedestrians.

\section{Related Work}

\subsection{Efficacy of VR Pedestrian Simulators}

There has been a notable increase in the utilisation of VR simulators for pedestrian behavioural studies in AV contexts~\cite{tran2021review, rouchitsas2019external, nascimento2019role}. These simulators may be classified according to their display systems, ranging from monitor-based systems~\cite{hoggenmuller2022designing}, through Cave Automatic Virtual Environment (CAVE)\hl{~\cite{madigan2023pedestrian, tian2023pedestrian, yang2024interpreting, tabone2024immersive}} to head-mounted displays (HMDs)\hl{~\cite{feldstein2016ped, sween2017development, deb2017efficacy, schneider2020virtually, bindschadel2021interaction}}. Another classification dimension is the nature of their environmental representation, which includes 360-degree real-world captures and computer-generated environments\hl{~\cite{hoggenmuller2019enhancing, hoggenmuller2021context, wang2024immersive}}. This paper focuses on fully immersive HMDs, which are increasingly preferred as the display type for VR pedestrian simulators\hl{~\cite{schneider2020virtually, tran2021review}}, and on computer-generated environments that allow for sophisticated configuration and facilitate interactivity within the virtual setting\hl{~\cite{tran2021review, hoggenmuller2021context}}.

The efficacy of a VR pedestrian simulator hinges on how closely pedestrians' perceptions and behaviour in virtual environments match those in real-world settings. Thus, the validation of VR for pedestrian safety studies necessitates evaluating the sense of presence pedestrians experience in the virtual environment, and ascertaining whether their crossing behaviour in VR mirrors that in the real world. In the validation study by Deb et al.~\cite{deb2017efficacy}, subjective feedback frequently reflected participants' perception of VR settings as believable and engaging. In contrast, objective data (including walking speed, crossing time, collisions, and the minimum gap between participant and vehicle) varied significantly under different traffic situations, demonstrating consistency with real-world behaviour.

Another validation study conducted by Bhagavathula et al.~\cite{bhagavathula2018reality} compared participants' performances in a real environment with those in a VR replica and in a 360-degree video recording of that environment. Participants were instructed to observe an approaching vehicle, and the assessment metrics encompassed their crossing intentions, estimations of the vehicle's speed and distance, as well as their perceived safety and risk. The results demonstrated negligible differences between real and virtual settings for the majority of metrics. However, disparities were noted in the virtual environment, particularly in the estimation of vehicle speed, which was higher in VR compared to the real world. Research by Feldstein et al.~\cite{feldstein2020road} illuminated more marked contrasts in participant behaviour between real and simulated settings. In real situations, pedestrians typically made their crossing decisions based on the temporal distance to the approaching vehicle, known as time-to-contact. In contrast, in the virtual milieu, crossing decisions appeared to depend more on the spatial distance from the vehicle, with less consideration for the vehicle's velocity.

Building upon these findings, our research introduces a novel perspective by incorporating multi-entity scenarios. The focus of our examination is on pedestrians' sense of presence in VR and their perceived similarity in crossing behaviours between VR and the real world. The latter aspect is evaluated through interviews with participants, in which they reflected on their crossing experience. This approach shifts the emphasis from objective measures (as in~\cite{deb2017efficacy, bhagavathula2018reality, feldstein2020road}) to the subjective perceptions and experiences of the participants. This shift is crucial, as comparing real-life and VR pedestrian crossing behaviours in multi-entity scenarios poses significant challenges, particularly in terms of the complex and potentially hazardous setup in the real world. Individual reflections, on the other hand, allow us to capture the nuances of human behaviours and identify specific factors that influenced their behaviour in simulated environments, such as perceived risks, environmental cues, and simulator realism.

\subsection{Considerations for VR Pedestrian Simulators} \label{considerations}

The pursuit of more effective VR simulations is pivotal in AV–pedestrian research. These efforts are directed towards improving the simulated experience, thereby yielding more accurate evaluations of pedestrian behaviour and their interactions with AVs. A number of studies have extrapolated considerations related to VR pedestrian simulators from thorough reviews of existing research on AV–pedestrian interactions~\cite{colley2019better, le2020automotive, tran2021review}. Most considerations focus on ensuring user health and comfort, exemplified by the use of the Simulator Sickness Questionnaire~\cite{kennedy1993simulator} to monitor simulation sickness or limiting session duration to under one hour~\cite{colley2019better, le2020automotive}. Immersion and presence are also paramount; for example, the incorporation of familiar elements, local traffic culture, background noise, and social atmosphere to enhance realism~\cite{tran2021review, le2020automotive}. Additionally, there are data collection considerations, such as integrating questionnaires within the VR environment to avoid an abrupt shift in context~\cite{schwind2019presence} or using a penalty-reward system to enhance its validity~\cite{hock2018how}.

The scalability issues of eHMI have increasingly come to the fore in recent years\hl{~\cite{colley2020unveiling, tran2023scoping, bazilinskyy2024not}}, necessitating the deployment of multi-entity VR simulations for the testing and analysis of more complex interactions. These simulations often encompass complex scenarios, such as navigating multi-lane roads amidst heavy traffic~\cite{mahadevan2019mixed, tran2022designing}, or crossing the street in the presence of other pedestrians\hl{~\cite{dey2021towards, mahadevan2019mixed, colley2022effects, lanzer2023interaction}}. These settings raise questions about the unique aspects of these interactions, especially in contrast to single-entity scenarios. A key consideration is how effectively VR can represent them. Due to the limited number of multi-entity VR simulations, employing a survey method to synthesise insights and practical considerations is not feasible. Moreover, an inherent limitation of review studies is the reliance on information provided by the included papers regarding the effectiveness of the simulations, in which there is often a lack of detailed reporting. In many cases, the assessment of participants' sense of presence was not conducted or reported. For example, in a review of 31 VR simulation studies by Tran et al.~\cite{tran2021review}, only 13\% reported measuring presence, and 39\% measured motion sickness. Therefore, our research employs a retrospective analysis of two simulations we developed. This firsthand involvement provides us with in-depth practical knowledge and understanding of the challenges and nuances involved.

\section{Method}

Our analysis was based on two studies examining pedestrian interactions with AVs in different virtual settings: \hl{\cite{tran2024exploring} and~\cite{tran2024evaluating}}. In developing the simulations for these studies, we adhered to the best practice recommendations for VR pedestrian simulator development where applicable (Section~\ref{considerations}). While both studies explored various concepts of AV external communication (i.e., eHMIs), the specifics of these eHMI designs and functionalities, although relevant, are beyond the scope of this paper and will not be discussed in detail.

The first study featured multi-vehicle environments, while the second centred on multi-pedestrian scenarios. The simulations in these two studies shared similar settings, such as 3D assets and VR hardware (collectively referred to as the `Shared Apparatus'), but were designed for distinct purposes. The multi-vehicle simulation placed more emphasis on traffic aspects, such as vehicle types, gaps between vehicles, and yielding behaviour in both lanes. In contrast, the multi-pedestrian simulation featured co-located pedestrians controlled by real humans and supported full-body avatar representation. As a result, utilising these two case studies provides a more comprehensive view by offering insights from two different angles. \autoref{tab:chap9-study-comparison} shows a summary of the participants' demographics and data collection in each study.

\begin{table}[t]
    \centering
    \small
    \caption{A summary of the participants' demographics and the data pertaining to VR experience, as collected in both case studies.}
    \begin{tabular}{lp{4cm}p{4cm}}
    \toprule
    & \textbf{Multi-Vehicle} & \textbf{Multi-Pedestrian} \\
    \midrule
    \textbf{N (m/f)} & 32 (16/16) & 25 (12/12, 1 undisclosed) \\
    \textbf{Age Groups} & & \\
    18-24 & 9 (28\%) & 6 (24\%) \\
    25-34 & 19 (60\%) & 17 (68\%)\\
    Above 34 & 4 (12\%) & 2 (8\%) \\
    \textbf{VR Familiarity} & & \\
    Never & 5 (16\%) & 2 (8\%) \\ 
    Less than 5 times & 13 (41\%) & 14 (56\%) \\ 
    More than 5 times & 14 (43\%) & 9 (36\%) \\
    \midrule
    \textbf{Data Collection} & & \\
    Sickness Questionnaire & Misery Scale & Misery Scale \\
    Presence Questionnaire & iGroup Presence Questionnaire & Multimodal Presence Scale \\
    Video Recording & First-person view & Third-person view \\
    Interviews & VR experiences, \newline Crossing behaviour & VR experiences, \newline Crossing behaviour, \newline Avatar perception \\
    \bottomrule
    \end{tabular}
    \label{tab:chap9-study-comparison}
\end{table}

\subsection{\hl{Apparatus and System Performance}}

\subsubsection{Shared Apparatus}

This section details the equipment and environment setup utilised.

\begin{figure}[t]
  \centering
  \includegraphics[width=0.75\linewidth]{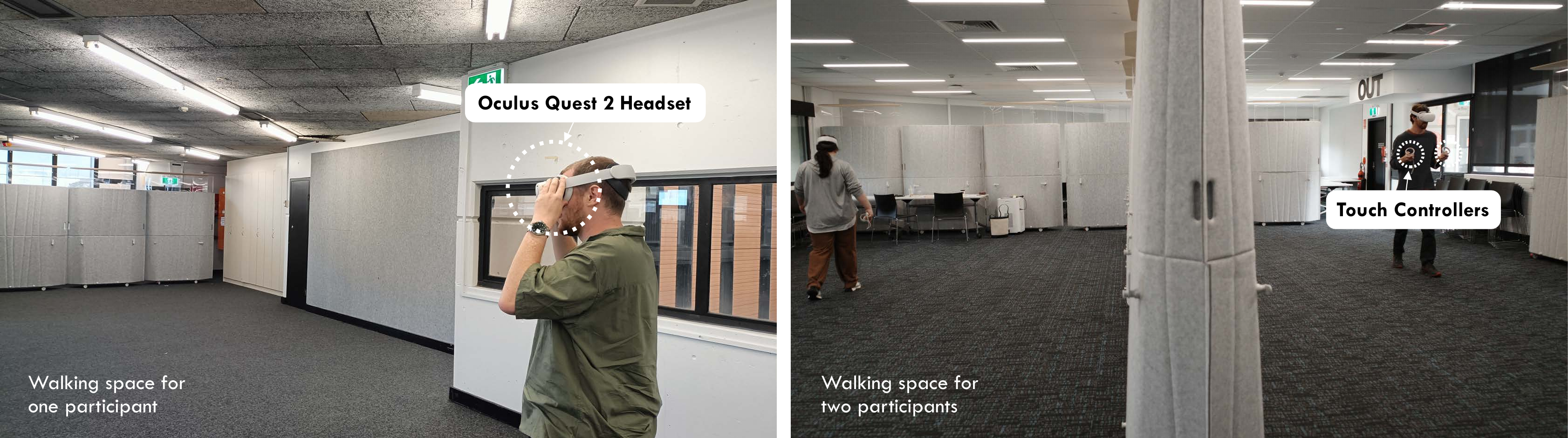}
  \caption{Physical space used in the multi-vehicle study (left) and in the multi-pedestrian study (right).}
  \label{fig:chap9-studysetup}
\end{figure}

\begin{itemize}
    \item \textit{Hardware}: We utilised the Meta Quest 2, a standalone VR headset. This device operates independently of a tethered PC, ideal for simulations requiring dynamic movement, such as pedestrian crossing. It weighs approximately 503 grams, and has a screen resolution of 1832 x 1920 pixels per eye. 
    \item \textit{Physical space}: \hl{The experiments were conducted in a studio, with an allocated walkable space measuring 8 m x 5 m, providing sufficient room for participants to navigate a two-lane street, with each lane measuring 3.5 m. A buffer was maintained around the walkable area for safety} (see \autoref{fig:chap9-studysetup}).
    \item \textit{Urban environment}: The virtual setting was designed using the Unity 3D game engine, incorporating assets from the Unity Asset Store. We specifically chose the Modern City Pack\footnote{\url{https://assetstore.unity.com/packages/3d/environments/urban/modern-city-pack-18005}} for its diverse and highly detailed building and street props, along with an ambient soundscape of birds and traffic to enhance realism.
    \item \textit{Participant criteria}: Participants were adults over 18 years old, with normal or corrected-to-normal vision. Those with colour blindness or mobility impairments that could influence the walking experience were excluded.
    \item \textit{Familiarisation session}: Participants were introduced to the VR equipment and spent time getting accustomed to the virtual environment. They practised crossing the street in VR, crossing at least twice in a simulated environment without vehicles.
    \end{itemize}

\subsubsection{\hl{System Performance}}

\hl{We used the OVR Metrics Tool\footnote{\url{https://developers.meta.com/horizon/documentation/unity/ts-ovrmetricstool}} to measure the frame rate of each simulation running on the Quest 2. The frame rate fluctuated depending on the complexity of the scene. During the familiarisation phase, which featured a static scene, and in the multi-pedestrian scenarios with only one vehicle approaching, the frame rate consistently remained between 60-72 Hz. In contrast, during the more dynamic multi-vehicle scenarios, with traffic on both sides of a two-way street, the frame rate averaged around 55 Hz and occasionally dropped to 50 Hz, particularly during longer sessions.}

\hl{Latency is typically a greater concern in multi-pedestrian simulations with more users or networked interactions. However, since our study involved only two users in the same physical space, latency was not an issue during either the pilot or official sessions. This conclusion is supported by two key observations: (1) the visual updates in the VR headsets were in sync with the users' actions, with no noticeable delays in rendering, and (2) the audio cues signalling the start of the scenario were synchronised across all devices.}

\subsection{Multi-Vehicle Simulation}

\subsubsection{Overview} This study compares interconnected eHMIs~\cite{tran2024exploring}, where external interfaces across different AVs synchronise to provide unified crossing information, with unconnected eHMIs operating independently. Focusing on pedestrian safety and multi-lane crossing experiences, the within-subjects design also includes scenarios without eHMIs.

\begin{figure}[ht]
  \centering
  \includegraphics[width=0.75\linewidth]{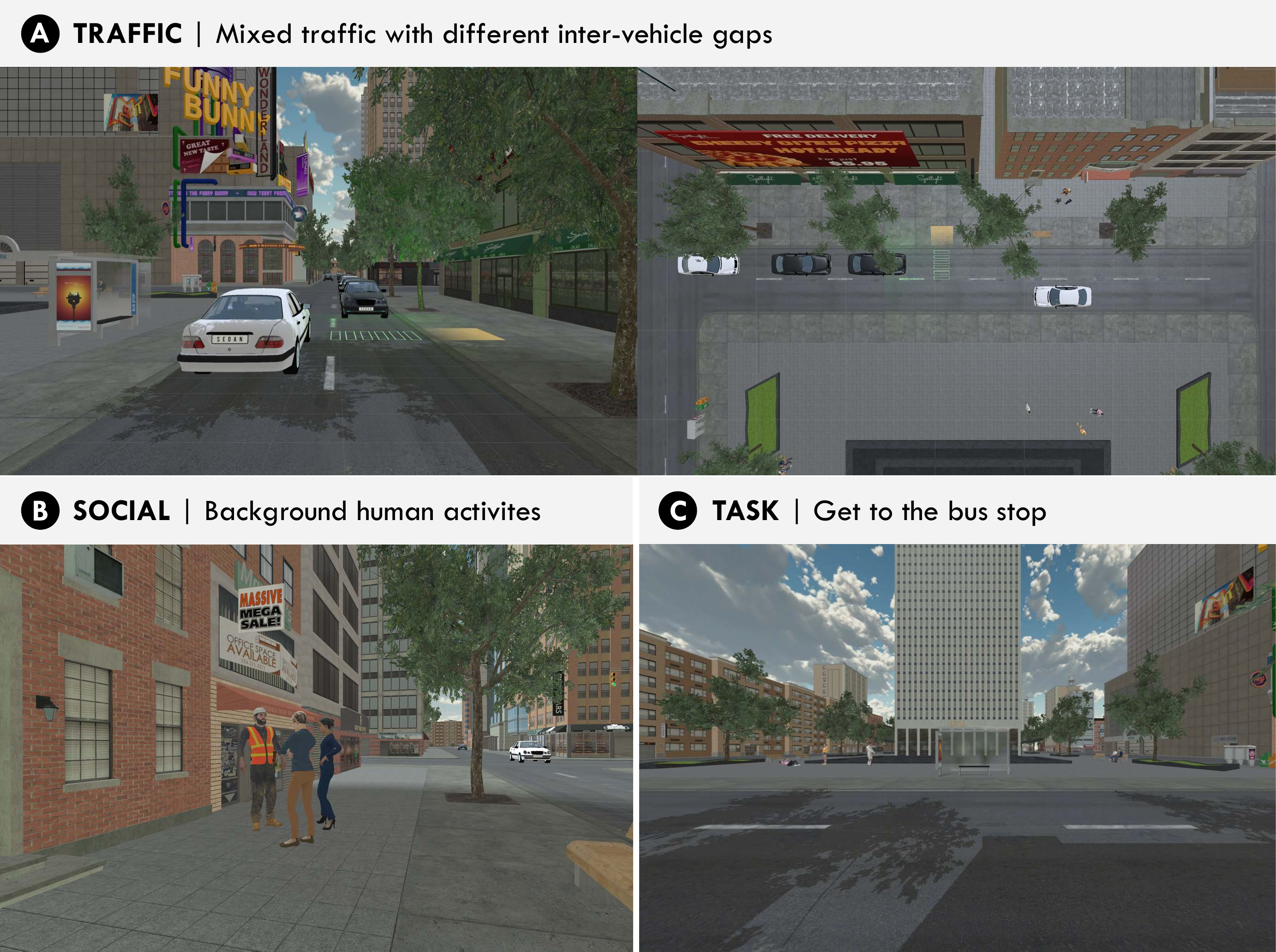}
  \caption{Settings of the multi-vehicle simulation: (A) Mixed traffic with AVs and manually-driven vehicles; (B) Human activities in the background; (C) A mid-block crossing where participants were tasked with reaching the bus stop.}
  \label{fig:chap9-multi-vehicle-setting}
\end{figure}

\subsubsection{Traffic Setting} 
The experiment used two types of vehicles: black sedans represented AVs without human occupants (SAE Level 5~\cite{sae2021taxonomy}), and white sedans symbolised manually operated vehicles with visible drivers. These vehicles were introduced in a mixed sequence, with inter-vehicle gaps varying from 2.5 to 6.5 seconds to mimic natural traffic flow (see \autoref{fig:chap9-multi-vehicle-setting}A). They emerged from a point beyond the participants' initial line of sight, \hl{with an initial speed of}
50 km/h. In scenarios where AVs are programmed to yield, the vehicles decelerated at 3.5 $m/s^2$ from a distance of 50 m, stopping 3 m away from the participant, adhering to recommended safety guidelines~\cite{qldgov2023}.

\subsubsection{Social Setting} In the simulation, the study participants had no virtual embodiment (avatar) as they were the only human pedestrians present, and prior research on pedestrian crossing behaviour found that a virtual body does not lead to increased presence~\cite{koojiman2019how}. However, the environment was populated with \hl{eight} Mixamo 3D characters, \hl{some standing in pairs and talking, while others were exercising in small groups,} to create a socially vibrant atmosphere~\cite{tran2021review} (see \autoref{fig:chap9-multi-vehicle-setting}B).

\subsubsection{Crossing Scenarios} The virtual environment featured a two-lane urban road with a midblock crossing. We created three distinct crossing scenarios. In each, an AV in the lane nearest to the pedestrian was programmed to yield 13 seconds into the VR scenario. The behaviour of vehicles in the far lane varied: either an AV stopped, a manually driven vehicle stopped, or no vehicle stopped. Participants were tasked with crossing the street to reach a bus stop on the opposite side (see \autoref{fig:chap9-multi-vehicle-setting}C). They completed nine crossing trials in total, spending approximately 10–12 minutes in the virtual environment.

\subsubsection{Data Collection} 
This section outlines the measures take to assess the VR simulation, excluding those related to eHMIs.

\begin{itemize}
   \item \textit{Sickness}: Following their experience with each eHMI concept in three scenarios, participants responded to the Misery Scale~\cite{bos2010effect} to monitor simulator sickness. The study was paused if ratings exceeded three to ensure participant well-being.
    \item \textit{Sense of presence}: After experiencing all eHMI concepts, participants completed the iGroup Presence Questionnaire (IPQ)\footnote{\url{https://www.igroup.org/pq/ipq/index.php}}~\cite{schubert2003sense}. This 14-item presence scale consists of four components: presence, spatial presence, involvement, and experienced realism.
    \item \textit{Semi-structured interviews}: Participants were then interviewed about their comfort with the VR headset, their impressions of the virtual environment, and whether their crossing behaviour in VR reflected real-world experiences.
    \item \textit{Video recording}: First-person perspective video data were captured using the Cast feature of the Meta Quest mobile application, documenting the participants' viewpoints and interactions within the VR simulation.
\end{itemize}

\subsection{Multi-Pedestrian Simulation}

\subsubsection{Overview} This VR study assesses communication between AVs and individual pedestrians\hl{~\cite{tran2024evaluating}}.
The two participants were co-located but not in a group setting, to avoid group influence~\cite{rasouli2019autonomous, nicolas2023social, rosenbloom2009crossing}. It compares three eHMI concepts: on the \textit{vehicle}, within the surrounding \textit{infrastructure}, and on the \textit{pedestrian} themselves. These are compared against a baseline condition without any eHMI, to assess the clarity of communication to the intended pedestrians.

\subsubsection{Traffic Setting} A black sedan AV is shown approaching from the left at a speed of 50 km/h (see \autoref{fig:chap9-multi-pedestrian-setting}A). When yielding to pedestrians, it decelerates at a rate of 2.4 $m/s^2$ at a 40-m distance (a normal braking rate based on previous research~\cite{ackermans2020effects, deligianni2017analyzing}).

\subsubsection{Social Setting} 
The simulation involved two human pedestrians interacting in real time with the AV\hl{, with no computer-controlled agents present}. Participants were represented by full-body avatars, pre-selected from Ready Player Me\footnote{\url{https://readyplayer.me/}}. The avatars' appearance was not personalised for several reasons: in the first-person view of the simulation, pedestrians were only able to see parts of their avatar, such as hands or feet, and not their full virtual representation. Additionally, only strangers were paired together in the same study session. Compared to existing approaches using pedestrian agent models~\cite{dey2021towards, mahadevan2019mixed, colley2022effects, lanzer2023interaction}, the use of a multi-pedestrian simulator is expected to enhance the social presence in VR scenarios~\cite{makransky2017development}, and simulate more realistic interaction experiences. The experiment utilised Meta Quest 2 VR headsets, providing precise tracking of head and hand movements. This tracking was complemented by inverse kinematics~\cite{aristidou2018inverse}, which estimate the motion of untracked body parts like elbows and legs.

The VR simulation was designed with multi-user networking capabilities, using the Photon Unity Networking framework\footnote{\url{https://www.photonengine.com/pun/}}. This setup allowed participants to interact within the same virtual space. In the experiment, participants were assigned one of two roles and alternated between them in different scenarios. The first pedestrian role placed the participant closer to the approaching AVs. The second pedestrian role was positioned further away, offering a consistent view of the first pedestrian's interactions with the AVs (see \autoref{fig:chap9-multi-pedestrian-setting}B).

\subsubsection{Crossing Scenario} 
\begin{figure}[t]
  \centering
  \includegraphics[width=0.75\linewidth]{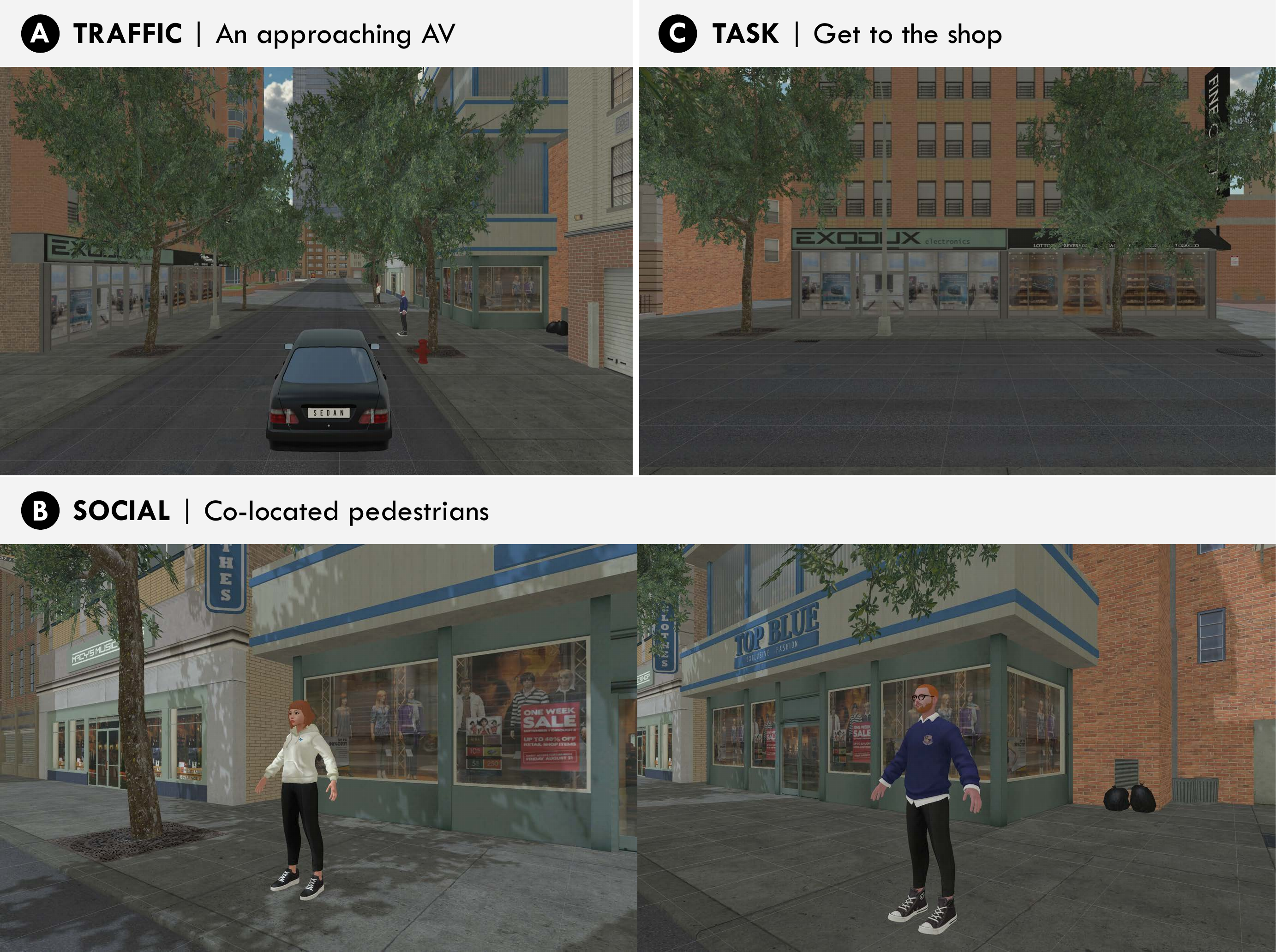}
  \caption{Settings of the multi-pedestrian simulation: (A) Traffic consists of one AV approaching from the left-hand side; (B) Two pedestrians crossing in the scenario; (C) A mid-block crossing where participants were tasked with reaching the shop in front of them.}
  \label{fig:chap9-multi-pedestrian-setting}
\end{figure}

The simulated environment replicated a two-way urban street, intentionally designed without traffic signals or pedestrian crossings. Two participants were situated 10~m apart on a pavement. Their task was to observe the approaching AV and the environment, then cross the street when deemed safe (see \autoref{fig:chap9-multi-pedestrian-setting}C). There were five scenarios, wherein the AV could stop for either of the two participants or not stop. Participants made a total of 20 crossing trials, spending approximately 20–25 minutes in VR.

\subsubsection{Data Collection}

\begin{itemize}
\item \textit{Sickness:} After experiencing each eHMI concept in three scenarios, participants responded to the Misery Scale~\cite{bos2010effect} to assess simulator sickness, pausing the study for ratings over three. 
\item \textit{Sense of presence}: Upon finishing their experience with all eHMI concepts, participants completed the Multimodal Presence Scale (MPS)~\cite{makransky2017development} to evaluate immersion and realism. This questionnaire was chosen as it features item groups focusing separately on physical, social, and self-presence, enabling a detailed assessment of the multi-pedestrian setup.
\item \textit{Semi-structured interviews}: The subsequent interviews delved into participants' perceptions of the virtual environment, the realism of their crossing behaviour, the influence of having a full-body avatar, and their opinions on the avatar's movement realism. 
\item \textit{Video recording}: Recordings were captured from a third-person perspective to facilitate external observation of both participants' movements and interactions in the VR environment. This perspective was achieved by including an invisible user in the same scenarios as the participants, who maintained a fixed viewing angle. A 2020 iMac with a 3.8 GHz 8-core Intel Core i7 processor was utilised for this recording process.
\end{itemize}

\section{Data Analysis}
Questionnaire data (Misery Scale, IPQ/MPS) were analysed to derive insights about participants' simulator sickness and sense of presence. Additionally, we explored the influence of participants' prior VR experiences on presence perceptions. \hl{Due to the unequal and small sizes of some groups, nonparametric statistical tests (Kruskal-Wallis H-test~\cite{kruskal1952use}) were chosen as they do not require the assumption of normality and are suitable for comparing medians across independent groups. We employed IBM SPSS Statistics Version 30.0.0.0 for all statistical analyses.}

Transcripts were subjected to thematic analysis~\cite{braun2006using}. We inductively identified recurring themes that shed light on the efficacy and potential shortcomings of each VR simulation. Video recording was used to review some notable crossing instances, corroborating the participants' responses.

\section{Results}

\subsection{Simulator Sickness}

In the multi-vehicle study, 29 out of 32 participants (91\%) averaged below 1.0 across all conditions, indicating no symptoms or mild discomfort. Three participants had a higher score (averaging 2.44), with some symptoms such as dizziness and headache. Similarly, in the multi-pedestrian study, 22 out of 25 participants (88\%) averaged below 1.0 across all conditions, with three participants having a higher score (averaging 1.75). In both studies, no participants had to suspend the study due to severe sickness symptoms.

\subsection{Presence Questionnaires}

\begin{figure*}[t]
  \centering
  \includegraphics[width=1\linewidth]{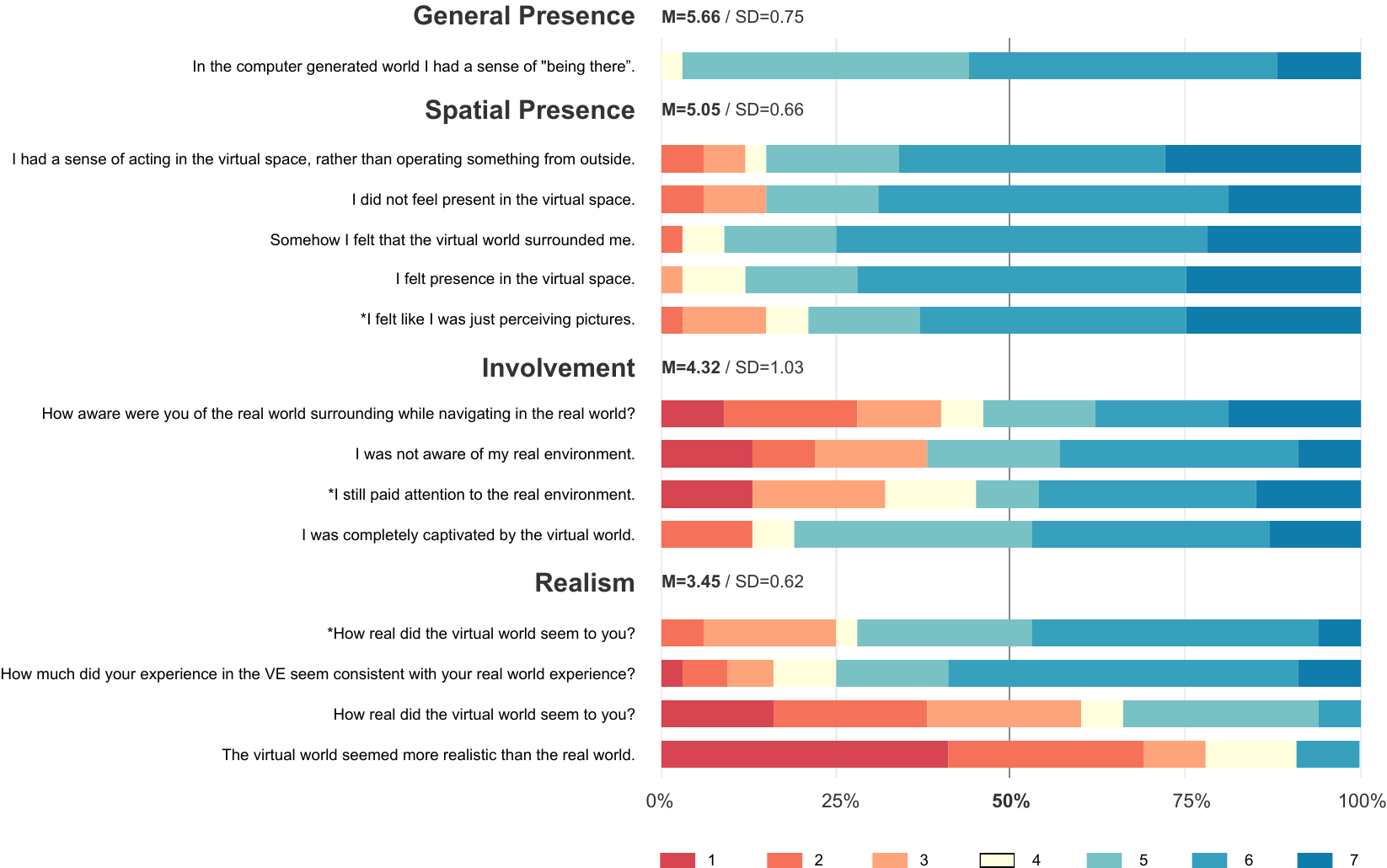}
  \caption{Visualisation of the responses to the iGroup Presence Questionnaire across various dimensions of presence: Spatial Presence, Involvement, Realism, and General Presence. Statements that were reverse-scored are indicated with an asterisk `*'. A higher score signifies a greater sense of presence.}
  \label{fig:chap9-IPQchart}
\end{figure*}

\subsubsection{Multi-Vehicle Simulation} 

To simplify interpretation, we converted the original range of IPQ Presence scores from -3 to 3 into a positive only range from 1 to 7. The overall IPQ Presence score for our 32 participants was 4.43 (SD~$=0.5$), with scores ranging from a low of 3.29 to a high of 5.36. 

Regarding IPQ subscales, the mean and distribution of responses (\autoref{fig:chap9-IPQchart}) suggests that the VR simulations are effective in creating a sense of presence, particularly in terms of \textit{General Presence} ({M~$=5.66$, SD~$=0.75$}), \textit{Spatial Presence} ({M~$=5.05$, SD~$=0.66$}), and \textit{Involvement} ({M~$=4.32$, SD~$=1.03$}). \hl{\textit{General Presence} refers to the overall feeling of `being there' in the computer-generated world, while \textit{Spatial Presence} describes the sensation of feeling present and actively engaging within the virtual space, rather than merely observing it. In contrast, \textit{Involvement} measures the degree to which users are engaged with the content and how much their attention is captured by the experience. Although participants felt strongly present within the virtual environment, their engagement was somewhat diminished due to external factors such as the physical conditions of the room.}

Participants' perceptions of \textit{Realism} ({M~$=3.45$, SD~$=0.62$}) \hl{were notably lower. While participants felt present within the VR environment, this did not translate into perceiving the environment as realistic}. One of the IPQ statements, \textit{`The virtual world seemed more realistic than the real world'}, is an extreme comparison, likely to prompt disagreement since most VR experiences do not yet surpass the fidelity of real-world experiences. Thus, lower scores in this area are expected and consistent with current technological limitations.


\hl{\textit{Impact of prior VR experience}}: 
Participants without VR experience and frequent VR users (`More than 5 times') both scored slightly below the overall sample average of 4.43 (M~$=4.37$, SD~$=0.28$ and M~$=4.35$, SD~$=0.56$ respectively). In contrast, individuals with a moderate level of VR exposure (`Less than 5 times') exhibited a slightly higher mean score (M~$=4.53$, SD~$=0.51$). \hl{A Kruskal-Wallis H-test revealed no statistically significant differences in IPQ scores across the VR familiarity groups, H(2) = 0.83, p = 0.660.}

\subsubsection{Multi-Pedestrian Simulation} 
\begin{figure*}[t]
  \centering
  \includegraphics[width=1\linewidth]{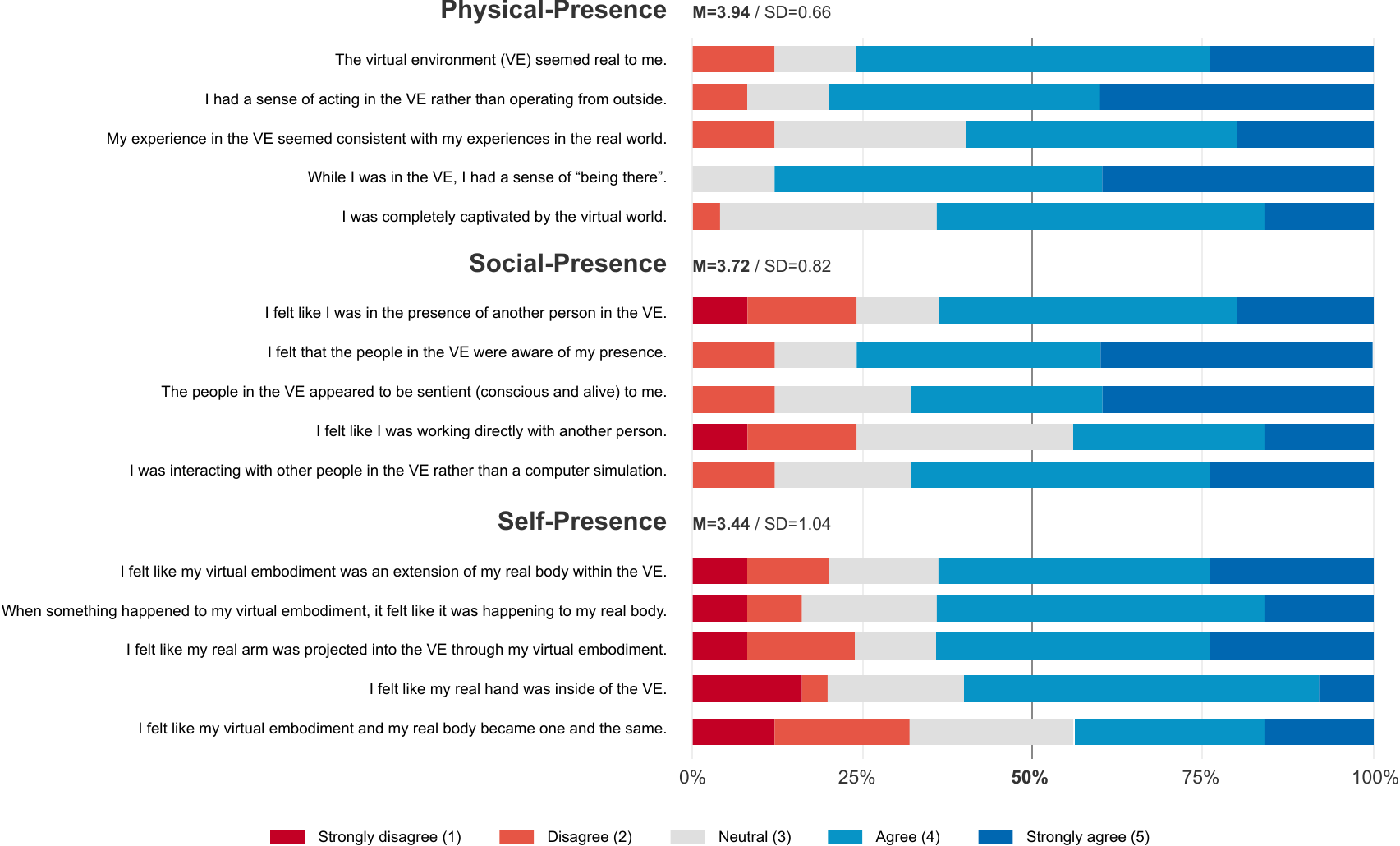}
  \caption{Visualisation of the responses to the Multimodal Presence Scale across various dimensions of presence: Physical Presence, Social Presence, and Self-Presence. A higher score signifies a greater sense of presence.}
  \label{fig:chap9-MPSchart}
\end{figure*}

There is no aggregated score for the MPS. Instead, it is broken down into subscales\hl{, with each item in the subscales rated on a 1 to 5 scale.} \textit{Physical Presence} leads with the highest average rating of 3.94 and the lowest variability (SD~$=0.66$). \textit{Social Presence}, while slightly lower in its average rating at 3.72, shows a slight increase in variability (SD~$=0.82$). The most noticeable variation is observed in \textit{Self Presence}, which not only has the lowest average rating of 3.44 but also the highest variability among participants (SD~$=1.04$). \autoref{fig:chap9-MPSchart} shows a more polarised set of responses for \textit{Self Presence}, with significant numbers in both the `Strongly agree' and `Strongly disagree' categories. Such a split could indicate that the simulation was highly effective for some in providing a sense of self-presence, while it was not effective for others.

\hl{\textit{Impact of prior VR experience}}: 
Two individuals with no VR experience reported notably elevated presence levels across all categories—Physical (M~$=4.6$), Social (M~$=4.7$), and Self (M~$=4.5$)—with exceptionally low standard deviations (SD~$\leq0.28$). 
Participants with a moderate level of VR experience displayed mixed results. Physical presence was rated moderately high (M~$=3.96$, SD~$=0.74$), but a noticeable decline was observed in both social (M~$=3.56$, SD~$=0.95$) and self presence (M~$=3.33$, SD~$=1.09$). Meanwhile, participants with extensive VR experience, reported the lowest presence levels in all categories—Physical (M~$=3.78$, SD~$=0.52$), Social (M~$=3.76$, SD~$=0.54$), and Self (M~$=3.38$, SD~$=1.00$). 
\hl{A Kruskal-Wallis H-test showed no statistically significant differences across the VR familiarity groups for the MPS subscales: Physical, H(2) = 2.47, p = 0.291; Social, H(2) = 2.67, p = 0.264; and Self, H(2) = 3.66, p = 0.160.}

\subsection{Interviews}

The structure of our thematic analysis is devised to unravel the layers of VR experience that influence participants' perception, starting with the \textit{physical} contexts, moving through the \textit{social} interactions, and culminating with the individual's sense of \textit{self}, all of which happens to mirror the structure of the MPS scale. Other important organising themes regarding crossing behaviour in VR and interaction with AVs were also identified. For ease of reference, we use \textcolor{teal}{\textbf{v}} and \textcolor{violet}{\textbf{p}} to denote participant quotes from multi-vehicle or multi-pedestrian simulations, respectively.

\subsubsection{Physical}

The analysis revealed that participants perceived a high degree of realism, noting the clear and consistent visuals, and the authentic representation of buildings and streets (\textcolor{teal}{v=15}, \textcolor{violet}{p=15}). Additionally, the soundscape, enriched with spatial audio effects from passing vehicles and ambient environmental noises, provided an immersive layer that contributed to the sense of being in a tangible space (\textcolor{teal}{v=12}, \textcolor{violet}{p=2}). The analysis suggests that although participants recognised the VR simulation was not a perfect replica of reality, its interactive nature requiring movement and observation significantly enhanced engagement (\textcolor{teal}{v=11}, \textcolor{violet}{p=9}). One reflected, \textit{`The active nature of this scenario, constantly moving and turning my head, made me feel more engaged'} (V20). Another participant noted, \textit{`I felt fully immersed in the environment, though I still could distinguish between the virtual reality and the real world'} (V17). 

Feedback on the speed of vehicles in the multi-vehicle simulation was mixed. While some participants felt the speed was realistic or appropriate (\textcolor{teal}{v=11}), others perceived it as too fast and suggested that the speed profile could be improved for realism (\textcolor{teal}{v=7}). Participants frequently mentioned difficulties in accurately estimating the speed of the vehicles and the safe gap for crossing (\textcolor{teal}{v=8}). Some participants experienced a discrepancy in the sense of scale within the virtual environment (\textcolor{teal}{v=4}). They felt smaller and noted that distances seemed greater, making tasks like crossing the road feel longer than in real life. These errors in judgment led to miscalculations and, in some cases, virtual collisions. 

Participants noted issues with the clarity and resolution of the VR experience, indicating that objects appeared blurry or low-resolution (\textcolor{teal}{v=9}, \textcolor{violet}{p=4}). This made it harder to see as clearly as in real life. There were comments about the unnatural textures of certain elements, such as trees. Additionally, some participants experienced graphical glitches, including jitteriness, lag, and choppiness, particularly when moving or turning their heads. 

\subsubsection{Social}

In the multi-vehicle simulation, participants responded favourably to the simulation of social activities within the virtual environment, appreciating how it added a sense of normalcy and vitality (\textcolor{teal}{v=12}). The presence of agents engaging in everyday behaviours, like conversing and exercising, contributed to a more vibrant and lifelike atmosphere. As V32 reflected, \textit{`It just made me feel like I'm not alone in that space. Although they are not real people, it just gives me a little bit more human connection.'} However, some participants pointed out that the simulation could be improved by addressing the busyness of the environment. V1 and V30, among others, mentioned that the agents' repetitive movements and the absence of a bustling crowd atmosphere sometimes detracted from the realism.

In the multi-pedestrian simulation, participants' feedback highlighted the positive impact of noticing the presence of other human participants (\textcolor{violet}{p=8}). The sense of authenticity was reinforced when participants observed gestures like waving, which prompted them to respond in kind, enhancing the feeling of being in a shared space. The ability to see and interact with other avatars was frequently mentioned as a key aspect of making the VR experience feel more social and lifelike. Participants repeatedly referred to the realistic movement of avatars, as noted by P1, \textit{`It seems very intuitive to me to see when someone is crossing.'}

In contrast, participants frequently noticed unnatural movements of others' avatars, which they described as weird or stuck (\textcolor{violet}{p=14}). The movements were sometimes so off-putting that it made the avatars seem like robots or not real people. One participant expressed initial uncertainty about whether the avatar they were interacting with represented a real person. Nevertheless, a number of participants indicated that despite noticing imperfections in the avatars' movements, it did not influence their decision-making process in a crossing scenario.

Of note, several instances of collisions occurred in the study. Despite observing avatars in distress or involved in accidents, the responses ranged from a complete lack of emotion to amusement (\textcolor{violet}{p=3}). Participants understood the simulated nature of the environment, which translated into a more detached and less empathetic reaction to the avatars' experiences, treating collisions as non-consequential or even as humorous occurrences, rather than distressing or alarming.

\subsubsection{Self}
 
Without an avatar in the multi-vehicle simulation, several participants noted that not being able to see their legs detracted from the experience (\textcolor{teal}{v=2}). P3 mentioned the need for clear visibility of their virtual body to judge their position correctly in relation to the virtual environment, such as standing at a crosswalk. V14 expressed discomfort due to losing track of their limbs and lacking a clear representation of their position in 3D space, which is crucial for spatial awareness.

Regarding the perception and embodiment of avatars in the multi-pedestrian simulation, several participants mentioned seeing only parts of their bodies, leading them to feel that the avatar had a limited impact on their experience (\textcolor{violet}{p=9}). The most commented upon visible part were the hands (\textcolor{violet}{p=12}). Participants frequently mentioned the added realism from being able to see and control their avatars' hand movements, such as waving. However, there were observations that the avatars' hands lacked realism, appearing as distinctly defined 3D models. Furthermore, the position and movements of the avatar could sometimes feel unnatural or distracting, as if its limbs and body parts were not properly aligned with the user's actual body. This dissonance could be jarring; one participant noted that their avatar's movements didn't feel like those of their real arms. In contrast to the hands, the aspect of seeing one's own feet was less commonly noted (\textcolor{violet}{p=2}). P20 remarked on the increased realism felt when they observed their shoes during walking, but P23 experienced a disruption in their sense of self when they could not see their feet clearly, indicating a reliance on foot positioning for spatial orientation within the VR environment.

\subsubsection{Crossing Behaviour}

Participants generally reported crossing streets in the VR environment in a manner similar to how they would in real life, where they relied on habitual safety considerations and traffic assessment (\textcolor{teal}{v=26}, \textcolor{violet}{p=20}). The VR simulation also evoked real emotions in participants, such as fear or caution, particularly when facing potential collisions or observing virtual car crashes (\textcolor{teal}{v=8}). However, the level of perceived risk and emotional response in VR differed from real life. Participants were conscious of risks but did not experience the same intensity of fear or urgency, as the VR environment lacked real physical harm. This discrepancy led to more relaxed behaviours and a reduced focus on potential dangers (\textcolor{teal}{v=6}, \textcolor{violet}{p=1}). The study's real world setting influenced participants' crossing behaviours in VR in the opposite way: they were more cautious and less inclined to move quickly due to the fear of running into physical objects in the real environments (\textcolor{teal}{v=6}, \textcolor{violet}{p=2}).

The tasks within the VR simulation, like waiting at a pedestrian crossing to reach a bus stop, were recognised for their realism and relevance to real-life scenarios, enhancing the authenticity of the experience (\textcolor{teal}{v=2}). However, the analysis captures an increased level of caution among participants, influenced by their consciousness of participating in an experiment (\textcolor{teal}{v=1}, \textcolor{violet}{p=3}). This cognisance appears to make them wary of how their riskier actions, which might be more typical in real life, could potentially alter the experiment's outcomes. 

In the context of multi-lane street crossing, it was evident that once participants commenced the crossing task, they rarely considered cancelling their crossing (\textcolor{teal}{v=2}). V28 remarked, \textit{`I was too focused on completing the crossing. Looking back, if I had the option, I would step back and not cross.} Likewise, V6 conveyed the cognitive intensity of the task, stating, \textit{`Because it was mentally demanding, I was solely focused on crossing the street and forgot about it [cancel crossing]'.} These insights underscore how the immersive quality of VR can markedly sway pedestrian decision-making behaviour.

\subsubsection{Expectations or Assumptions about AV Behaviour} \label{expectations}

In both studies, the yielding behaviour of the AV was preprogrammed. Depending on scenario, the AV was set to yield or not yield to the participants, and this response was fixed, regardless of their intentions or behaviour. Without knowing this, participants experienced uncertainty or confusion regarding how AVs work, specifically related to aspects such as pedestrian detection, intention recognition and decision-making. Their feedback revealed certain expectations or assumptions about AV behaviour: 

(1) Participants expected the AV to respond to their behaviour: Several participants deliberately tested the limits and capabilities of the vehicle by moving closer to the edge of the road, presumably with the expectation that the AV would detect them more easily, thus increasing the likelihood of the AV stopping for them (\textcolor{teal}{v=6}, \textcolor{violet}{p=1}). One participant noted, \textit{`I wasn't sure what triggered the stops. I think it was when I was close enough to the lane and took a couple of steps forward'} (V24). In these instances, pedestrians sometimes used gestures like waving as a probing action to communicate with or ascertain the AV's intentions.

(2) Due to these expectations, participants appeared confused when the AV seemed unresponsive to their actions, highlighting uncertainties about the AV's decision-making process (\textcolor{teal}{v=7}, \textcolor{violet}{p=15}). In scenarios with multiple vehicles, the primary uncertainty was which vehicle would yield, and in situations with multiple pedestrians, the question was about which pedestrian the AV would allow to proceed.

This is exemplified when a participant remarks, \textit{`I thought the car was going to stop, but it just kept going,'} (V17) and another adds, \textit{`It was unclear why the car stopped, and that was disconcerting'} (V18). These statements underscore the unpredictability of AV behaviour from a pedestrian's perspective. Furthermore, many participants questioned the reliability of AVs when the vehicle failed to yield to them. As V26 stated, \textit{`Well, some autonomous cars didn't recognise me and just drove right by.'} These statements indicate that participants linked a vehicle's decision to yield with its ability to recognise pedestrians. When an AV fails to stop, these participants are inclined to assume that the vehicle failed to recognise them. This assumption is especially strong in situations with multiple pedestrians, where an AV might stop for one person but not another. P4 described this as feeling almost discriminatory, \textit{`Why did the vehicle stop for someone else and not for me? Maybe the autonomous vehicle didn't know I was there. It makes me think there might be a malfunction.'} Only in some rare cases was the influence of the vehicle's owner on its decision to yield suggested: \textit{`I don't think AV would be driven solely by the urgency of the situation, but rather the attitude and preferences of the owner'} (V18). 

\section{Discussion}

In this section, we reflect on the findings related to the factors that enhance or diminish the sense of presence for participants in AV–pedestrian simulations, aiming to derive recommendations for improving VR pedestrian simulators, and suggest areas for future research.

\subsection{Presence Dimensions in AV–Pedestrian Simulations}

The presence questionnaires from both simulations showed that although the virtual world's realism was not perceived as hyperrealistic and was greatly influenced by real-world factors, the simulations were effective in creating a moderate sense of presence. This success was primarily attributed to the compelling visuals and soundscapes, as well as the interactive nature of the simulations.

\textit{Visual aspects}: While the visuals in virtual environments were appreciated for their consistency and authenticity, achieving higher levels of realism remains a complex goal. This challenge is partly due to the uncanny valley effect~\cite{mori2012uncanny}, where near-realistic human figures or animations can evoke feelings of eeriness or discomfort among users~\cite{hoggenmuller2021context}. Additionally, the limitations in processing power and graphical rendering capabilities can impede the creation of highly detailed and realistic environments. Despite these challenges, advancements in VR technology, including the development of digital twins of cities—virtual replicas of existing districts or cities~\cite{BloombergCities2021}—promise significant improvements in visual fidelity and familiarity. It is important to note, however, that VR simulations, while valuable, cannot yet fully replace real-world testing. Their current role is primarily as a preliminary tool in design, offering initial insights before more comprehensive real-world evaluations.

\textit{Soundscapes}: The integration of sound in VR was found to be one of the critical components for making the environment feel more believable. Investigating the impact of real-world ambient audio recordings and synthesised virtual AV sounds on presence, Dongas et al.~\cite{dongas2023virtual} highlighted the significance of ambient soundscapes and the spatialisation of vehicle sounds to add plausibility and depth to the virtual environment. However, a review of VR studies on AV–pedestrian interaction\cite{tran2021review} observed that the majority of studies tend to neglect the sound aspect, including vehicle engine noise and ambient sounds. This omission can detract from the immersive quality of the simulation, suggesting that a balanced audio-visual approach is essential for creating a convincing virtual environment.

\textit{Interactivity}: In line with existing research, the interactivity offered by computer-generated environments is a key factor in their effectiveness. This advantage is highlighted in the comparison with hyperreal prototyping methods that use 360-degree video recordings~\cite{dongas2023virtual, hoggenmuller2021context}. While the latter may offer higher visual fidelity, computer-generated environments excel in user engagement, allowing participants to be actively involved rather than just passive observers.

In the context of these evolving capabilities and constraints, Dongas et al.~\cite{dongas2023virtual} have contributed to the discussion by positing that the dimensions of presence to maximise in a VR simulation are contingent upon its intended use and context. This viewpoint is particularly relevant to our studies, where we focused on enhancing physical presence in multi-vehicle simulations, and social and self presence in multi-pedestrian simulations. The following sections will discuss the results and implications of these specific dimensions.

\subsubsection{Multi-Vehicle Simulation}

\paragraph{\textbf{Addressing Challenges in Multi-Lane Street Crossing}} In multi-vehicle scenarios, where participants must cross multi-lane streets, it is crucial to accurately simulate the spatial and temporal dynamics of moving vehicles. This precision is essential for participants to correctly judge the distance and time required to safely cross the street. However, qualitative feedback indicates that VR simulations face challenges in this regard. Issues like a limited field of view (FOV) and blurry visuals hinder the ability to accurately estimate these parameters. Similar concerns were also noted in a previous study focusing on mixed traffic simulations~\cite{mahadevan2019mixed}. To mitigate these hardware limitations, using advanced VR headsets can be beneficial. For example, compared to the Meta Quest 2 used in our study, the Meta Quest 3 offers a wider FOV (110 degrees) and a higher resolution (2064 x 2208 pixels per eye), representing an approximately 30\% improvement. \hl{To maintain smooth frame rates in scenes with heavy traffic, optimisation techniques such as occlusion culling should be implemented to reduce rendering demands.}

In the study, when crossing in the virtual environment, participants often chose not to move quickly or run due to fear of colliding with real objects. Although using an untethered headset in a larger space helped, concerns about the real environment still affected their experience, as shown by low \textit{Involvement} scores on the IPQ presence questionnaire. A few participants also felt the virtual world's scale was off, making them feel smaller and the street longer to cross. In retrospect, these issues might have been better addressed with a comprehensive familiarisation session focusing on street crossing at various speeds and including height calibration.

\paragraph{\textbf{Considering Self-Avatar Integration}} In our multi-vehicle simulation study, participants were not provided with an avatar. During interviews, only a few comments indicated the potential need for an avatar to achieve a more effective representation in 3D environments. This suggests that while avatars might not strongly influence participants' sense of presence, they could potentially improve tasks such as multi-lane road crossing, where participants sometimes briefly wait in the middle of the road. 

The use of avatars in VR pedestrian simulators is still a relatively uncharted area, with only a handful of studies exploring this aspect. These studies often employ advanced tracking systems like motion capture suits~\cite{koojiman2019how, feldstein2016ped}, which use reflective markers for precise tracking of numerous body joints. The primary aim of using avatars in these experiments is to enhance the sense of presence for participants in VR. However, a study by Koojiman et al.~\cite{koojiman2019how} found no significant improvement in participants' subjective experiences using this method. Meanwhile, participants in the study by Feldstein et al.~\cite{feldstein2016ped} found avatars to be valuable, but they considered the graphical quality to be subpar. In both instances, the lack of realism was identified as a key issue. The low adoption rate of avatars can be attributed to hardware requirements, the complexity of setup~\cite{feldstein2016ped, schneider2020virtually}, and the lack of solid evidence regarding the relevance of an avatar in simulations that lack social factors (such as those involving a single participant). Despite these challenges, we advocate for further research into embodiment in VR simulations, particularly those involving multi-lane scenarios, to enhance presence and foster more realistic crossing behaviour.

\subsubsection{Multi-Pedestrian Simulation}

\paragraph{\textbf{Simulating More Accurate Leg Movements}} Findings suggest that participants initially focused intently on visible body parts, mirroring observations from a simulation by Feldstein et al.~\cite{feldstein2016ped}. The qualitative feedback emphasised that embodiment in VR significantly enhanced participants' engagement with the virtual environment, making it feel like an extension of their physical reality. However, they also pointed out areas needing improvement. Specifically, they noted the need for better synchronisation between real and virtual hand movements and expressed a desire for the ability to bend down and view their legs in the virtual space. This limitation highlights a common challenge in VR HMDs, which is the accurate tracking and representation of the entire body, especially the legs. Without an advanced tracking system, approaches utilising inverse kinematics (implemented in our simulation) are occasionally accurate for elbows and rarely correctly for legs. Meta AI researchers have developed a new approach called Avatars Grow Legs (AGRoL), which utilises a diffusion model for real-time body pose estimation. However, this technology is not yet widely available for consumer VR products~\cite{du2023legs}.

In the context of multi-pedestrian simulations, the issue of leg tracking and smooth movement also negatively impacted social presence. In rare cases, these issues led participants to question the authenticity of their virtual co-participants. While participants indicated that these realism deficiencies in movement did not affect their crossing decisions, we propose that they might still obscure subtler aspects of body language, such as slight hesitations, minor step-backs, or subtle leaning motions. Future research should consider comparing various implementation strategies. For instance, Feng et al.~\cite{feng2023another} adopted the use of legless avatars in their multi-pedestrian simulations, which might offer different insights into the impact of avatar realism on user experience.

\paragraph{\textbf{Reconsidering the Role of Avatar Self-Similarity}} An interesting finding was that most participants were indifferent to whether the avatars resembled themselves. This echoed a finding from our pilot test and was the main reason behind our decision not to provide each participant with a customised avatar that resembled their own physical appearance. A recent study by Kim et al.~\cite{kim2023tobe} found that a high level of avatar self-similarity boosts users’ sense of embodiment and social presence, yet it has minimal effects on overall presence and may even slightly hinder immersion. Additionally, in our multi-pedestrian simulation, the social context involves interactions between two strangers; therefore, identity might not play a major role.

\paragraph{\textbf{Meeting Experienced Users' Expectations}} Descriptive analysis indicated a potential relationship between participants' familiarity with VR and their perceived sense of presence in the multi-pedestrian simulations. The higher sense of presence among new VR users may be attributed to the novelty effect, where unfamiliar experiences are perceived as more intense or engaging. In contrast, the lower sense of presence among more experienced users suggests that participants with prior VR familiarity may approach the experience with a more evaluative stance. This finding aligns with existing literature, suggesting that individual user characteristics, such as previous experience with VR and expectations regarding the simulated experience, are among the variables that influence the feeling of presence~\cite{coelho2006media, kalawsky1993comprehensive}. 

In the multi-vehicle simulation, however, we observed a more consistent sense of presence across both experienced and inexperienced users. We posit that the pronounced interpersonal interactions and avatar/embodiment issues in the multi-pedestrian simulation lead experienced VR users to be more critical and evaluative. While further investigation is needed to fully grasp how VR experience impacts presence in different simulation contexts, our findings underscore the importance of refining social and self presence elements in VR simulations. This refinement is crucial for ensuring a strong sense of presence, particularly in simulations with complex social dynamics.

\subsection{The Influence of Controlled Simulation}

The heightened sense of responsibility and self-consciousness observed among participants, influenced by their awareness of being part of a study, aligns with the Hawthorne effect. This phenomenon, widely documented in behavioural studies, suggests that individuals modify their behaviour in response to their awareness of being observed~\cite{mccarney2007hawthorne}. Conversely, some participants exhibited riskier behaviour due to the perception of being in a safe, simulated environment. This finding is in line with the theory of risk compensation, which posits that individuals adjust their risk-taking behaviours based on their perception of safety~\cite{wilde1982risk}. As a result, it becomes pertinent to explore alternative assessment methods that could mitigate these influences. One such approach is the use of stealth assessment~\cite{shute2011stealth, young2024ishot}, which subtly disguises the primary goal from the participants. For example, instead of explicitly instructing participants to cross a street in the presence of AVs, they could be tasked with navigating a simulated cityscape where encounters with AVs occur naturally. This method not only integrates the task more seamlessly into the environment, but it also reduces the participants' focus on being observed or assessed, potentially yielding more authentic behavioural responses. The reward-penalty system~\cite{hock2018how, colley2019better} offers a promising approach to eliciting more genuine behaviour from participants. By rewarding safe interactions with AVs and penalising risky actions, the simulation may encourage natural responses, enhancing the authenticity of data on decision-making processes. This idea, though not yet widely implemented in any VR pedestrian simulators, aligns with gamification principles, using game-like incentives~\cite{almeida2024analysis, chang2024must} to boost engagement and improve outcomes. 

\begin{figure}[t]
  \centering
  \includegraphics[width=0.60\linewidth]{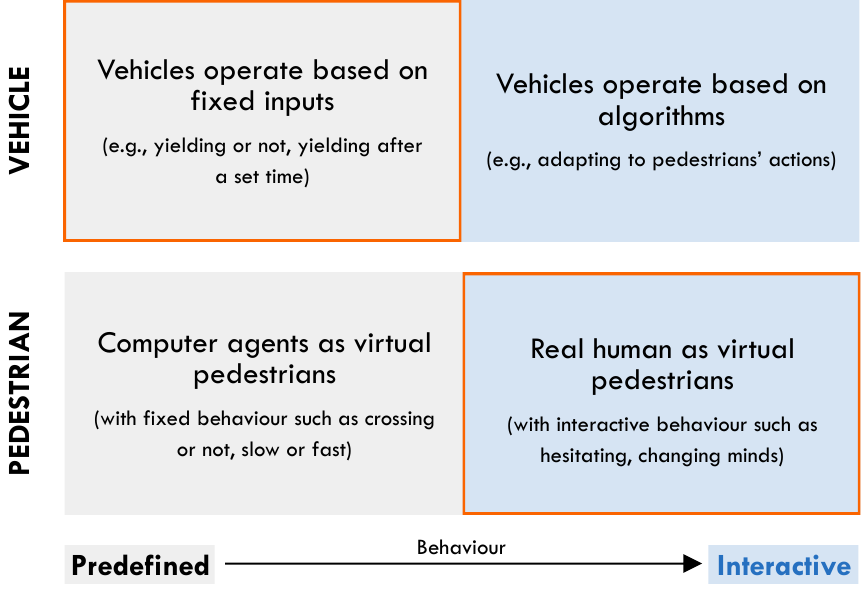}
  \caption{The quadrant diagram presents a comparison between predefined and interactive responses of vehicles and pedestrians within the context of VR simulations. \hl{The orange box highlights the approach used in this study's simulations.}}
  \label{fig:agent_matrix}
\end{figure}

The absence of bidirectional interaction due to the predefined AV behaviour in simulations represents another limitation, particularly when contrasted with the dynamic and unpredictable nature of actual AV–pedestrian interactions. This discrepancy becomes apparent in instances such as those documented in publicly shared videos of Waymo and Tesla Full Self-Driving cars~\cite{brown2023halting}. In one example, a man, accompanied by his family, signalled an oncoming car to proceed, yet the vehicle slowed down instead. The situation became more complex when his wife started crossing and the car unexpectedly accelerated, prompting an apology from the AV's driver: \textit{`Sorry, it's a self-driving car'}. The lack of interactive elements in the simulation makes the simulations less representative of real-world scenarios, introducing uncertainty or confusion regarding how AV works (as in Section~\ref{expectations}). Tran et al.~\cite{tran2021review}'s review also noted the scarcity of interactive AV behaviour in VR simulations, with Camara et al.~\cite{camara2020examining} providing one of the few examples where an AV's movement was adaptively altered based on pedestrian positions. \hl{As shown in \autoref{fig:agent_matrix}, our simulation has achieved more interactive pedestrian behaviour; however, future research should focus on incorporating more responsive and adaptive AV behaviours to better reflect real-world scenarios.}

\medskip

\subsection{Limitations}
A notable limitation of this study stems from the potential disruption in presence (known as `break-in-presence' (BIP)~\cite{jerald2017vrbook}). This disruption is caused by participants completing the Presence questionnaires only after they have removed their VR headsets. While filling out questionnaires within the VR setting does not appear to affect presence measurements, enhancing response consistency has been noted~\cite{schwind2019presence}. Moreover, the study's design involved initially interviewing participants about eHMI concepts. Discussions about the VR simulations occurred only at the study's conclusion, introducing a time gap between the VR experience and the subsequent interviews. This gap could potentially affect the accuracy and immediacy of the participants' responses regarding the VR simulations. To address this issue, we are planning a follow-up study. In this study, we will use micro-phenomenological interviews\footnote{\url{https://www.microphenomenology.com/interview}} to more closely examine participants' experiences within the virtual environment~\cite{knibbe2018dream, prpa2020articulating}.

\section{Conclusion}
The retrospective analysis of this study on two VR-based traffic scenarios, encompassing multiple vehicles and pedestrians, has provided valuable insights into the effectiveness of multi-entity VR simulations. Our findings emphasise the importance of enhancing participants' sense of presence to ensure realistic and immersive experiences in such simulations. It is noteworthy that different dimensions of presence should be given priority depending on whether the focus is on vehicular traffic or pedestrians.

The research underscores the substantial influence of controlled scenarios in VR on participants' crossing behaviour and their interactions with AVs. This observation, highlighting the need for more interactive and dynamic simulations, is pivotal for the future design of VR simulations. Such improvements could lead to more accurate and reliable data on pedestrian and AV interactions, contributing significantly to the development of safer and more efficient traffic systems.


\begin{acks}
This research is supported by an Australian Government Research Training Program (RTP) Scholarship and through the Australian Research Council (ARC) Discovery Project DP220102019, Shared-space interactions between people and autonomous vehicles. \hl{We thank Prof. Martin Tomitsch for supervising the two simulation studies, as well as all participants for their involvement. We also appreciate the valuable feedback provided by the anonymous reviewers.}
\end{acks}

\bibliographystyle{ACM-Reference-Format}
\bibliography{main}

\appendix

\end{document}